# An Optimized Huffman's Coding by the method of Grouping

Gautam.R
Department of Electronics and Communication Engineering,
Maharaja Institute of Technology, Mysore
gautamrbharadwaj@gmail.com

Dr. S Murali
Professor, Department of Computer Science Engineering
Maharaja Institute of Technology, Mysore
murali@mitmysore.in

*Abstract*— Data compression has become a necessity not only in the field of communication but also in various scientific experiments. The data that is being received is more and the processing time required has also become more. A significant change in the algorithms will help to optimize the processing speed. With the invention of Technologies like IoT and in technologies like Machine Learning there is a need to compress data. For example training an Artificial Neural Network requires a lot of data that should be processed and trained in small interval of time for which compression will be very helpful. There is a need to process the data faster and quicker. In this paper we present a method that reduces the data size. This method is known as Optimized Huffman's Coding. In the Huffman's coding we encode the messages so as to reduce the data and here in the optimized Huffman's coding we compress the data to a great extent which helps in various signal processing algorithms and has advantages in many applications. Here in this paper we have presented the Optimized Huffman's Coding method for Text compression. This method but has advantages over the normal Huffman's coding. This algorithm presented here says that every letter can be grouped together and encoded which not only reduces the size but also the Huffman's Tree data that is required for decoding, hence reducing the data size. This method has huge scientific applications.

*Keywords-Huffmans coding; Data Compression; Signal processsng;*

## I. INTRODUCTION

Data compression has become a necessity while processing information that occupies huge memory. Data compression not only reduces the size but also helps to increase the processing time. Huffman's coding is based on the frequency of occurrence of a data item. The principle is to use a lower number of bits to encode the data that occurs more frequently. Codes are stored in a code book. In all cases the code book plus encoded data must be transmitted to enable decoding.

Huffman coding is a lossless data compression algorithm. The main principle in Huffmans coding is that each message in encoded using individual or combination of 0 or 1 bits. Here the most frequent characters will be assigned with the smallest code and the least frequent character will have the largest code. Hence Huffman's coding is based on the variable length approach. Also in Huffman's there is no repetition of the code for the same message.

Here in the Optimized method for construction of minimum redundancy code algorithm letters are grouped into sets of 2 or 3 depending on the size of the data. Huffman's coding basically works on the principle of frequency of occurrence for each symbol or character in the input. For example we can know the number of times a letter has appeared in a text document by processing that particular document. After which we will assign a variable string to the letter that will represent the character. Here the encoding take place in the form of tree structure which will be explained in detail in the following paragraph where encoding takes place in the form of binary tree.

### A. *Huffmans Coding*

First, in this section we explain normal Huffman's coding. Huffman encoding is a way to assign binary codes to symbols that reduces the overall number of bits used to encode a typical string of those symbols. There are mainly two major parts in Huffman Coding: Build a Huffman Tree from input characters and traverse the Huffman Tree and assign codes to characters. Decoding is more or less the reverse process, based on the probabilities and the coded data, it outputs the decoded byte.

Steps for Huffman's coding. Create a leaf node for each unique character and build a group of all leaf nodes (Group is used as a priority queue. The value of frequency field is used to compare two nodes in group. Initially, the least frequent character is at root. Then we extract two nodes with minimum frequency from the Tree.

Later we create a new internal node with frequency equal to the sum of the two nodes frequencies. Make the first extracted node as its left child and the other extracted node as its right child. Add this node to the Tree. We repeat the process until the Tree contains only one node. The remaining node is the root node and the tree is complete.

To show this method of Huffman's coding, we have used text as an example. Consider the below Text example of Huffman's coding.

"IEEECOMPUTATIONALINTELLIGENCE"

Here in the above text we have 29 letters. The frequency and character is as shown below

TABLE 1

| Character | Frequency |
|---|---|
| E | 6 |
| I | 4 |
| T | 3 |
| N | 3 |
| L | 3 |
| C | 2 |
| O | 2 |
| A | 2 |
| M | 1 |
| P | 1 |
| U | 1 |
| G | 1 |

In the above TABLE 1 the characters are arranged in the descending order of their frequencies. The encoded data for the above data is as shown in the below table.

TABLE 3

| Character | Encoded Bits |
|---|---|
| E | 11 |
| I | 10 |
| T | 0111 |
| N | 0110 |
| L | 0101 |
| G | 0100 |
| O | 0011 |
| A | 0010 |
| M | 00011 |
| P | 00010 |
| U | 00001 |
| G | 00000 |

From the above TABLE 3 we can know that using Huffman's coding only 100 bits are required. If we don't use Huffman's coding then we need 232 bits (29 characters* 8 bits).

The decoding procedure is very simple. As the bits are are read, i.e when the first bit is read from the input, it is traversed from the beginning of the root by taking the left path or the right path in the binary tree depending on the the binary code 0 or 1 respectively. When the end of the tree is reached, characters would be decoded, and that character is placed on the output stream. So all the characters are decoded.

*B. Optimized Huffmans Coding by grouping them in the set of 2*

Below method shows the table (TABLE 2) of characters and their corresponding frequency. In this algorithm we group the letter into set of 2.

TABLE 2

| Character | Frequency |
|---|---|
| EI | 10 |
| TN | 6 |
| LC | 5 |
| OA | 4 |
| MP | 2 |
| UG | 2 |

When Huffman's coding is applied to the above TABLE 2 considering each of the two letter as one we get the encoded bits. Here now for two letters we get one encoded bits so we must split the letter into two. We do it by taking the first set, in this case EI and split it into E and me. Now we check for the frequency of the letter, so the letter which gets highest frequency will be directly encoded with the obtained bit and the other character will be encoded with the addition of 0. The table (TABLE 4) is shown below

TABLE 4

| Character | Encoded Bits |
|---|---|
| EI | 10 |
| TN | 100 |
| LC | 101 |
| OA | 110 |
| MP | 1110 |
| UG | 1111 |

Now that we have obtained the Encoding bits for the letters in the set of 2, we need to arrange them. As explained previously, in EI, E is having highest frequency so it will be encoded as 1 and I will be encoded with a prefix 0 i.e. 01. Here E is encoded as 1 while I is encoded as 01 and so on. So in this method we use only 82 bits to represent the above text.

## C. Optimized Huffmans Coding by grouping them in the set of three

We have seen grouping of Huffman's coding in the set of two, now we will look at grouping in the set of three.

Consider the below TABLE 5 of Huffman's coding grouped into set of three. In this algorithm we group the letter into set of 3 as show below.

TABLE 5

| Character | Frequency |
| --- | --- |
| EIT | 13 |
| NLC | 8 |
| OAM | 5 |
| PUG | 3 |

When Huffman's coding is applied to the above TABLE 5 considering each of the three letters as one we get the encoded bits as shown in the below TABLE 6.

TABLE 6

| Character | Encoded Bits |
| --- | --- |
| EIT | 0 |
| NLC | 10 |
| OAM | 110 |
| PUG | 111 |

To encode each bits we do it the same as that explained in the optimized Huffman's Coding for set of two. We find the letter with highest frequency and encode it. Here in this case, consider EIT where E has the highest frequency followed by I and then T. Now E is encoded directly with the obtained bits as 0, I is encoded with the prefix 0 i.e. the encoded bits is 00 and T is encoded with the prefix 1 to the obtained encoded bit i.e. 10 and so on. Here only 77 bits are sufficient to represent the given characters which shows an amazing compression rate.

## D. Algorithm

The algorithm can be well understood after the above examples

- First, we arrange the letters in descending order of the frequency
- Second, we group the letters in the set of 2 or 3 or 4 etc. accordingly and then we tabulate them.
- Third, we perform Huffman's coding to each of the letter sets.
- Fourth, we obtain the encoded bits.
- Next we see the frequency of the letter in each of the sets and arrange the letters in the specific set in descending order.
- Then we encode each letter. The letter with the highest frequency is encoded with the obtained bit directly. The following letter is encoded with the prefix 0, the following next letter with 1 and so on.
- To decode we perform normal decoding procedure followed in Huffman's coding.

.

## II. RESULTS

The results show that Optimized Huffman's coding can be implemented successfully. This approach helps research teams to make the results more efficient. This method is applied to text and the results are listed below. So the results are displayed categorically as:

1. UTF-8
2. Huffman's Coding
3. Optimized Huffman's Coding using set of 2
4. Optimized Huffman's Coding suing set of 3

Compression ratio is calculated with respect to that of UTF-8 coding. The results are listed below.

The Huffman's coding results and comparison is as show below in TABLE 7.

TABLE 7

| Name | Encoded Bits | Compression Ratio |
| --- | --- | --- |
| UTF-8 | 232 bits | 0% |
| Huffman's coding | 99 bits | 57.33% |
| Optimized Huffman's Coding using set of 2 | 82 bits | 63.37% |
| Optimized Huffman's Coding using set of 3 grouping | 77 bits | 76.30% |

## III CONCLUSION

The results obtained in this new method will be very helpful for many research works. It not only reduces the bits but also the tree because while decoding, the Huffman Tree is needed and in this method the Huffman tree is reduced to a great extent. Most of the researchers use huffmans coding in their algorithm and this optimized Hufmmans coding will help those researchers very much especially in signal processing applications and in various algorithms.

### ACKNOWLEDGMENT

This work was supported by Maharaja Institute of Technology and inTalks organization started in Maharaja Institute of Technology to promote research and innovation among students